\documentclass{iopconfser}

\newcommand{\HH}{{\mathbf{H}}}
\newcommand{\R}{{\mathbf{R}}}
 \newcommand{\SSS}{\mathbf{S}} 

\newcommand{\bq}{\begin{quote}}
\newcommand{\eq}{\end{quote}}

\newcommand{\be}{\begin{equation}}
\newcommand{\ee}{\end{equation}}

\newcommand{\ben}{\begin{enumerate}}
\newcommand{\een}{\end{enumerate}}
\newcommand{\bit}{\begin{itemize}}
\newcommand{\eit}{\end{itemize}}
\newcommand{\edoc}{\end{document}}

\newcommand{\gs}{g_{\SSS^{n-1}}}

\begin{document}

\title{Cosmological spacetimes with spatially constant sign-changing curvature}

\author{
Miguel S\'anchez 
}

\affil{
Departamento de Geometr\'{\i}a y Topolog\'{\i}a, Universidad de Granada, Granada, Spain.}

\email{sanchezm@ugr.es}

\begin{abstract}

Globally hyperbolic spacetimes endowed with a time function $t$ whose spacelike slices $t=t_0$ have constant curvature $k(t_0)$ and where the sign of $k(t_0)$ (as well as the topology of the slice) varies with $t_0$, can be constructed despite  some common claims about the implications of the classical  Cosmological Principle. Here, we  stress the possibilities of these cosmologies and announce the development of new models obtained in collaboration with  G. García-Moreno, B. Janssen, A. Jiménez-Cano, M.~Mars and R. Vera, \cite{GJJMSV}.\footnote{The very careful reading and constructive observations by the referee are deeply acknowledged. The support of the coauthors of \cite{GJJMSV}  is also warmly acknowledged, especially the detailed reading by A. Jim\'enez Cano and comments  by G. Garc\'{\i}a-Moreno, M. Mars and R. Vera. Work partially supported by the  project PID2024-156031NB-I00 and IMAG–Mar\'{\i}a de Maeztu grant CEX2020-
001105-M, both funded by Spanish MCIN/AEI/10.13039/50110001103.}
\end{abstract}

\section{A hidden cosmological possibility in the Cosmological Principle (CP)}

From a classical viewpoint (as, for example, \cite{CB}), 
the CP comprises two ingredients:

\ben\item Existence of a set of freely falling  fundamental observers (galaxies) 
which are  
synchronizable by using their proper time.  This leads to the {\em existence of a cosmic time}  $t$.

\item Considerations on observed  isotropy, which turns out in the requirement that {\em the space  at each instant $t=t_0$  has constant curvature} $k(t_0)$.
\een 
In the literature, sometimes
Friedmann-Lemaitre-Robertson-Walker (FLRW) cosmologies, here regarded as free of restrictions on their scale factors $a(t)$, are achieved 
 just from these  facts, say:

\bit\item The global manifold structure should be  a product $I\times \Sigma$, where $I\subset \R$ is an interval.
\item The metric should have the type $g^{(4)}=-dt^2+g^{(3)}_t$ where
  $g^{(3)}_t$ is a 
$t$-dependent metric of  constant curvature $k(t)$ on $\Sigma$.
\item 
  $g^{(3)}_t= R(t) g_\epsilon$, where $R(t) >0$ and, for a  unique  $\epsilon=1,0,-1$,
  
$
g_{\epsilon}:= \hbox{standard metric on}
\left\{
\begin{array}{llc}
\SSS ^3 \, \hbox{(3-sphere)} & \hbox{if} \;  \epsilon= & 1 \\
\R ^3 \,  \hbox{(3-Euclidean)} & \hbox{if} \; \epsilon= & 0 \\
\HH ^3 \, \hbox{(3-hyperbolic)} & \hbox{if} \; \epsilon= & -1, 
\end{array}
\right.
$
\eit 
the last point yielding the
 three types of standard FLRW cosmologies with no transitions,  
 but this is misleading. Indeed, as shown recently in
\cite{S23}, one can construct   3+1 spacetimes $(M,g^{(4)})$ satisfying:

\bit\item   $M= \cup_{t\in I} \left(\{t\}\times \Sigma_t\right)$, $t\in I\subset \R$ ($M$ is just foliated, thus, only a local product),

\item The metric is type   $g^{(4)}=-dt^2+g_t^{(3)}$, where   $g^{(3)}_t$ is a  
$t$-dependent metric of  constant curvature $k(t)$ 
 and both the sign of $k(t)$ and topology of $\Sigma_t$  change with $t$, and  

 \item No ``unfair tricks'' are used: they are  {\em smooth} and {\em globally hyperbolic. } 
\eit
That is,  spacetimes satisfying the CP (as formulated above) and  admitting transitions of sign in the curvature or topology are possible. We will refer to them as {\em counterexamples} to FLRW cosmologies.

These counterexamples can be disregarded by looking at the precise mathematical assumptions on the slices $t=t_0$ implying FLRW spacetimes. Namely,    
each point $p$ must  admit a neighborhood such that, given two directions 
$v, w$ at $p$ tangent to the  slice 
$t=t(p)$, there exists an observer-preserving isometry  which sends $v$ 
to $w$ 
(\cite[Ch 12, Prop. 6]{O},  
see also \cite[\S 5.1]{W} and the recent  revision of  cosmological principles in \cite{EMM}).  
 Nevertheless, one can wonder to what extent such mathematical hypotheses are physically meaningful (see \'Avalos' revision \cite{A23}  of the  physical notions on isotropy motivated after \cite{S23}).

However, the counterexamples yield new possibilities such as: 


\ben\item Topological transitions might permit us to 
match a finite Big-Bang with the  observed flatness of the Universe. Indeed, one might start at some time $t=t_0$ in a compact (Cauchy) hypersurface $\Sigma_{t_0}$ (thus with  finite 
energy and matter) and arrive at $t=t_1$ to a non-compact (and non-Cauchy) slice $\Sigma_{t_1}$ with Euclidean intrinsic  geometry. 


\item   The existence of   two different  types of time on the same spacetime emerges: 

\ben\item[(a)] The {\em Cauchy time} associated with  predictability. For this time, the space at each instant $t_0$ has a fixed topology, but no good properties of curvature or symmetry.
This time comes from a purely mathematical construction (see below), and  it would   
 be available only 
for  omniscent observers.

\item[(b)] The {\em (cosmic)   curvature time} directly associated with matter and energy and, thus, (in principle) measurable. Topologically more flexible, its compatibility with the Cauchy time may provide   links with  a sort of inflation  \cite{S23, GJJMSV}. 
\een

\een
This motivates a more 
in-depth  exploration of their relevant properties in the forthcoming work  \cite{GJJMSV}.
Here, we focus on global hyperbolicity, as this property (satisfied by all FLRW spacetimes) is philosophically appealing and underlines the CP. 

More precisely, let us start at stably causal spacetimes, which  can be defined as $(n+1)$-spacetimes $M$ admitting a time function  $t$ (that is, $t$ is  a continuous function and increases strictly along future-directed causal curves). Noticeably, one can then find a more restrictive {\em temporal} one, which is smooth   with gradient  $\nabla t$, past-directed and timelike \cite{BS05,S06}.  Such a function induces a global splitting   $TM=(\nabla t) \oplus $ Ker($dt$) and  local ones of type $\R\times S$ for $M$.  It also  provides the natural field of observers on $M$: 

 \begin{equation} \label{e_T}
 T:=-\nabla t/|\nabla t|.
 \end{equation}
Globally hyperbolic spacetimes can be defined as
being causal (that is, no causal loop exists\footnote{This weakens the classically  imposed condition of strong causality, see \cite{BS07}.}) 
with no naked singularities (which means that any $J(p,q):= J^+(p)\cap J^-(q)$ is compact).  
A celebrated theorem by Geroch \cite{G70} asserts 
that global hyperbolicity is equivalent to the existence of a (topological)  Cauchy hypersurface  $\Sigma$, leading to a global   
 topological splitting $M=\R\times \Sigma$. The fact that $\Sigma$ can be obtained  smooth and spacelike (see \cite{BS03}),   not only improves the global splitting into a smooth one but also links global hyperbolicity to the {\em  predictability} of the spacetime (no global obstruction appears for the   well-posedness of an initial value problem on $\Sigma$). 
Moreover, as proven in \cite{BS05}, globally hyperbolic spacetimes admit  temporal functions with Cauchy slices, so that the spacetime splits globally and orthogonally as:
   
 $$\R \times \Sigma, 
\, g= -\beta d\tau^2 + g_\tau   
 =  -\beta(\tau, x^k) d\tau^2 + \sum_{i,j=1}^n g_{ij}(\tau,x^k)dx^idx^j 
$$ 
Notice that the corresponding  observers in $T$ 
(as in (\ref{e_T})) 
are well adapted to the global 
causal structure  but, in general, the $\tau$-
slices do not satisfy any local property which 
could be measured by them. In FLRW spacetimes, the 
natural time satisfies   both local 
and global satisfactory  properties.

\section{Three counterexamples}
Our counterxamples are obtained starting at three different representations of FLRW spacetimes. Next, an open subset of each one is defined by using  points  $(t,x)$ in the product $I\times \R^n$.

\subsection{The $k(t)$-warped model {\em \cite{S23}}.} 
Let $\gs$ be the  metric of the sphere, $r$ the radial coordinate in $\R^n$. Choosing  $ t\mapsto k(t) $  
smooth,   
$$
 g_{war}= -dt^2 + dr^2 +S_{ k(t)}^2(r)   \, \gs \, ; \; \; 0< r <d_k= 
\left\{ \begin{array}{cl}
 \frac{\pi}{\sqrt{k}} & \hbox{if $k>0$}\\
 \infty &  \hbox{if $k\leq 0$}
\end{array} \right. ; \; \;  S_k(r)\ :=
\left\{ \begin{array}{ccl}
\frac{\sin(\sqrt{k}\, r)}{\sqrt{k}} & & \hbox{if} \; k>0 \\
r & & \hbox{if} \; k=0 \\
\frac{\sinh(\sqrt{-k}\, r)}{\sqrt{-k}} & & \hbox{if} \; k<0, 
\end{array}\right. $$
 is smooth, as so is the function $(t,k)\mapsto S_k(r)$ (indeed, analytic). Moreover, it extends smoothly at $r=0$ and it  admits a  small perturbation (preserving  constant curvature for $t$-slices) so that it can be also extended to  $r=d_k$, that is, the whole sphere, when $k(t)>0$.   
Choosing
$	k(t) >0$ if $ t<0$ and $	k(t) \leq 0 $ if 
$t \geq 0$,  a topological and curvature sign  change in the  $t$-slices occurs. However, the spacetime admits  compact Cauchy hypersurfaces. 
 Moreover, all the  choices of the function $k(t)$ for which the resulting
spacetimes are globally hyperbolic can be characterized, see  \cite{GJJMSV}.


\subsection{The $k(t)$-conformal model  {\em \cite{GJJMSV}}.} 
All the possibilities when  
	global hyperbolicity holds,  are also characterized for the spacetime:

$$g_{conf}=-dt^2 + \frac{1}{\left(1+\frac{ k(t) }{4} r^2\right)^2} \left(d r^2+  r^2 g_{\SSS^{n-1}}\right)$$ 
		
		$$ \hbox{defined in the open set} \qquad 
			U^{conf}= \left\{(t,x)\in I\times\R^n: 0<r(x) <\frac{2}{\sqrt{-k(t)}}  \;  (\hbox{when} \;  k(t)<0)  \right\},
		$$
		which is directly extended to $r=0$ and, whenever $k(t)>0$, to a whole sphere  $\SSS^n$ corresponding to the limit $r=\infty$. 	

\subsection{The $k(t)$-radial model {\em \cite{GJJMSV}}.} 
Our third model admits a spatial curvature sign change, but not a topological change:

$$g^{rad}=-dt^2 + \frac{1}{1- k(t) r^2} d r^2+  r^2 g_{\SSS^{n-1}}$$  

$$
U^{rad}=\left\{(t,x)\in I\times\R^n: 0<r(x)<1/\sqrt{k(t)}   \;  (\hbox{when} \;  k(t)>0)  \right\}.$$

In striking difference with the previous two models, when $k(t)>0$ the metric cannot be extended to the whole  sphere (now corresponding to the limit $r=1/\sqrt{k(t)}$ ), but  only  to an open half sphere. 
Then, the $k(t)$-model one  becomes  spatially open with no topology change in the $t$-slices (and trivially smooth).
Anyway, it is  globally hyperbolic for suitable choices of $k(t)$.

\bigskip


\noindent As a final remark, the scenarios that may lead to such a spatial curvature variation, along with their observational detectability, warrant further investigation. As pointed out in \cite{GJJMSV}, the matter content associated with the three novel geometries corresponds to a fluid exhibiting radial anisotropy. Their compatibility with an acceleratedly expanding universe that asymptotically approaches a Euclidean phase suggests that these models could be an interesting alternative to the standard FLRW models, although careful further analyses and comparison with observational data would be required.



\end{document}